\documentclass[journal,12pt,twoside]{IEEEtran}
\usepackage{cite}
\usepackage{acro}
\DeclareAcronym{SOP}{
short=SOP,
long= state of polarization,
}

\DeclareAcronym{AIR}{
short=AIR,
long= achievable information rate,
}

\DeclareAcronym{DSP}{
short=DSP,
long= digital signal processing,
}

\DeclareAcronym{PMD}{
short=PMD,
long= polarization-mode dispersion,
}

\DeclareAcronym{DOF}{
short=DOF,
long= degrees of freedom,
}

\DeclareAcronym{CMA}{
short=CMA,
long= constant modulus algorithm,
}
\DeclareAcronym{MMA}{
short=MMA,
long= multi-modulus algorithm,
}
\DeclareAcronym{RDE}{
short=RDE,
long= radially directed equalizer,
}

\DeclareAcronym{QAM}{
short=QAM,
long= quadrature amplitude modulation,
}

\DeclareAcronym{LMS}{
short=LMS,
long= least mean square,
}

\DeclareAcronym{LS}{
short=LS,
long= least-square error,
}

\DeclareAcronym{DP}{
short=DP,
long= dual-polarization,
}

\DeclareAcronym{PN}{
short=PN,
long= phase noise,
}

\DeclareAcronym{GN}{
short=GN,
long= Gaussian Noise,
}

\DeclareAcronym{NLI}{
short=NLI,
long= nonlinear impairments,
}

\DeclareAcronym{MIMO}{
short=MIMO,
long=  multiple-input multiple-output,
}

\DeclareAcronym{SNR}{
short=SNR,
long= signal-to-noise ratio,
}

\DeclareAcronym{MI}{
short=MI,
long= mutual information,
}

\DeclareAcronym{pdf}{
short=pdf,
long= probability density function,
}

\DeclareAcronym{AWGN}{
short=AWGN,
long= additive white Gaussian noise,
}

\DeclareAcronym{ASE}{
short=ASE,
long= amplified spontaneous emission,
}

\DeclareAcronym{QPSK}{
short=QPSK,
long= quadrature phase-shift keying,
}

\usepackage{hyperref} 
\usepackage{graphicx}
\graphicspath{../Figure/}
\DeclareGraphicsExtensions{.pdf,.jpeg,.png,.jpg, .svg}
\usepackage{xcolor}
\usepackage{tikz}
\usepackage{caption}
\usepackage{subcaption}
\usepackage{makecell}
\usepackage{psfrag}

\usepackage{amsmath}
\usepackage{amsfonts}
\usepackage{mathrsfs}
\usepackage{array}
\usepackage{theoremref}

\usepackage{url}

	\DeclareMathOperator{\tr}{tr}
	\DeclareMathAlphabet\mathbfcal{OMS}{cmsy}{b}{n}
	\definecolor{myMagneta}{rgb}{1, 0, 1}
    \newtheorem{theorem}{Theorem}
    \newcounter{lemma}
    \newtheorem{corollary}[lemma]{Corollary}
    \newtheorem{lemma}{Lemma}

	\renewcommand{\b}[1]{\mathbf{#1}}
	\newcommand{\rvec}[1]{\mathbf{\underline{#1}}}
	\renewcommand{\vec}[1]{\underline{#1}}

    \newcommand\norm[1]{\left\lVert#1\right\rVert}
    \newcommand{\I}{\mathrm{I}}
    \newcommand{\In}{\mathrm{I}_n}
    \newcommand{\E}{\mathbb{E}}
    \newcommand{\LambdaHat}{\hat{\Lambda}_{\vec{x}}}
    \newcommand{\Lambdax}[1]{\Lambda_{\vec{#1}}}
    \newcommand{\pin}{\pi^n}
    \newcommand{\Q}{\mathrm{Q}}
    \renewcommand{\H}{\mathrm{H}}
    \newcommand{\Hu}{\mathrm{H_u}}
    \newcommand{\Hb}{\mathbf{H}}
    \newcommand{\Hbu}{\mathbf{H}_\mathrm{u}}
    \newcommand{\Hhat}{\mathrm{\hat{H}}}
    
    \newcommand{\Hbhat}{\mathbf{\hat{H}}}
    
    \newcommand{\sigmaz}{\sigma_{z}^2}
    \newcommand{\sigmazn}{\sigma_{z}^{2n}}
    \newcommand{\CN}{\mathcal{CN}}
    \newcommand{\RE}{\mathrm{R}_\mathrm{E}}
    \newcommand{\logFrac}[2]{\log\left(\frac{#1}{#2}\right)}
    \newcommand{\expFrac}[2]{\exp\left(-\frac{#1}{#2}\right)}
\begin{document}

\title{
Capacity Bounds under Imperfect Polarization Tracking
}


\author{Mohammad~Farsi,~\IEEEmembership{Student Member,~IEEE,} Magnus~Karlsson,~\IEEEmembership{Senior Member,~IEEE} and Erik~Agrell,~\IEEEmembership{Fellow,~IEEE,} 
\thanks{M. Farsi and E. Agrell are with the Department
of Electrical Engineering, Chalmers University of Technology, SE-41296 Gothenburg, Sweden (e-mail: farsim@chalmers.se; agrell@chalmers.se)}
\thanks{M. Karlsson is with the Department of Microtechnology
and Nanoscience, Chalmers University of Technology, SE-41296 Gothenburg,
Sweden (e-mail: magnus.karlsson@chalmers.se).}
\thanks{This work was supported by the Knut and Alice Wallenberg Foundation under grant 2018.0090. }}
%
\maketitle

\begin{abstract}
In optical fiber communication, due to the random variation of the environment, the state of polarization (SOP) fluctuates randomly with time leading to distortion and performance degradation. The memory-less SOP fluctuations can be regarded as a two-by-two random unitary matrix. In this paper, for what we believe to be the first time, the capacity of the polarization drift channel under an average power constraint with imperfect channel knowledge is characterized. An achievable information rate (AIR) is derived when imperfect channel knowledge is available and is shown to be highly dependent on the channel estimation technique. It is also shown that a tighter lower bound can be achieved when a unitary estimation of the channel is available. However, the conventional estimation algorithms do not guarantee a unitary channel estimation. Therefore, by considering the unitary constraint of the channel, a data-aided channel estimator based on the Kabsch algorithm is proposed, and its performance is numerically evaluated in terms of AIR.
Monte Carlo simulations show that Kabsch outperforms the least-square error algorithm. In particular, with complex, Gaussian inputs and eight pilot symbols per block, Kabsch improves the AIR by $0.2$ to $0.35$ bits/symbol throughout the range of studied signal-to-noise ratios.

\end{abstract}
\section{Introduction}
\IEEEPARstart{T}{he} growing demand for reliable long-distance communication makes it essential to determine the capacity limits of optical links \cite{Monroe2016,Agrell:2016,Essiambre:2010}. The emergence of coherent optical communication systems enabled \ac{DSP} and polarization-division multiplexing to achieve higher spectral efficiencies. However, higher data rates come with increased sensitivity to impairments such as \ac{PMD} and \ac{SOP} fluctuations, which must be tracked dynamically in the receiver \cite{Savory:2010}. Due to the random variation of both internal and environmental impairments, the \ac{SOP} fluctuates randomly with time \cite{Poole:1988}. 
Previous long-term measurements showed that the \ac{SOP} drift might vary from days and hours in buried fibers \cite{Karlsson:2000,Allen:2003} to microseconds in aerial fibers \cite{Waddy:2001,Krummrich:2016,Zhang:2006}. 

The conventional \ac{DSP} solutions for \ac{SOP} tracking are replete with both blind and data-aided algorithms. The \ac{CMA} \cite{Godard:1980}, thanks to its low complexity and tolerance to \ac{PN}, has been widely considered for blind polarization tracking \cite{Savory:2007, Savory:2008, Kikuchi:2008}.
Many studies have been conducted to overcome the so-called ``singularity" problem of \ac{CMA} \cite{Kikuchi:2011,Liu:2009,Faruk:2010}, where the two estimated channels converge twice to the same polarization. 
The \ac{RDE} and its variants were proposed to account for higher-order \ac{QAM} in \cite{Ready_RDE:1990,Fatadin_RDE:2009,Lavery_RDE:2015}. The \ac{MMA} \cite{Yang_MMA:2002} is also applicable to higher-order modulations.
More reliable convergence can be obtained using data-aided algorithms such as \ac{LMS}, which was adopted for \ac{SOP} estimation by \cite{Savory:2008,Fludger_LMS:2008}, or standard \ac{LS}, which has been extensively studied for both wireless and optical applications \cite{Barhumi:2003,Ip:2007,Shieh:2008,Kuschnerov:2010}. 
While the \ac{LMS} algorithm performs real-time continuous equalization, \ac{LS} applies to block-based estimation.

While the \ac{DP} channel can be represented by a unitary matrix, the majority of the available estimation techniques including \ac{LS} and \ac{LMS} have no unitary constraint on the estimated channel, leading to sub-optimal solutions. This makes room for algorithms that provide a unitary estimation of the \ac{DP} channel. Louchet \textit{et al.} \cite{Louchet:2014} proposed the Kabsch algorithm \cite{Kabsch:1976} as a joint blind \ac{PN} and \ac{PMD} estimation method. In \cite{Czegledi:2016}, a blind modulation-format-independent joint \ac{PN} and polarization tracking algorithm was proposed, and its performance was compared with the blind Kabsch algorithm. 

Depending on the estimation algorithm, the speed of the fluctuations, or the additive noise, the polarization tracking might be imperfect. This makes it relevant to ask how the capacity is affected by such an estimation error? While there are no capacity studies regarding the polarization drift channels, to the best of our knowledge, the literature is replete with fiber capacity results. The first finite capacity limit for the single-wavelength optical channel based on a sort of \ac{GN} model was introduced in \cite{Splett:1994,Kurtzke:1995}, showing a lower bound that increases with the power until it reaches a maximum, and then it decreases due to the \ac{NLI}. Although different nonlinear models were applied, similar results were reported in \cite{Green:2002,Mitra:2002}. Since the \ac{GN} model cannot truly describe the \ac{NLI} \cite{Agrell:2014}, some modeled the \ac{NLI} as a linear time-variant distortion highly correlated over time \cite{Secondini:2012,Secondini:2013}. This made it possible to counteract \ac{NLI} using the techniques that were conventionally used for linear impairments (e.g., \ac{PN}, chromatic dispersion, \ac{PMD}, and \ac{SOP} \cite{Secondini:2014,Dar:2014_1,Dar:2015}). For example, in \cite{Dar_BlockWise:2014}, a tighter lower bound for single-polarization, dispersion-unmanaged systems was achieved by considering the \ac{PN}.

In this paper, for the first time, we study the capacity of the polarization drift channel in the presence of imperfect knowledge of the channel (e.g., imperfect polarization tracking).
Using the mismatched decoding method, we derive an AIR, which is a lower bound on the capacity, in the presence of a channel estimation error. We show that a unitary estimation of the channel leads to a tighter lower bound, which makes it reasonable to seek unitary estimators. A data-aided version of the Kabsch algorithm is proposed for \ac{DP} channel estimation, for the first time to the best of our knowledge. We compare Kabsch with an \ac{LS} algorithm in terms of \ac{AIR} and show that Kabsch outperforms \ac{LS} throughout the range of considered \acp{SNR}. For instance, with only eight pilot symbols per channel and block, Kabsch improves the \ac{AIR} by at least $0.2$ bits per symbol compared to \ac{LS}.

\textit{Notation:} Column vectors are denoted by underlined letters $\vec{x}$ and matrices by uppercase roman letters $\mathrm{X}$. We use bold-face letters $\mathbf{x}$ for random quantities and the corresponding nonbold letters $x$ for their realizations. \Acp{pdf} are denoted as $p_{\b{x}}(x)$ and conditional \acp{pdf} as $p_{\b{y|x}}(y|x)$, where the subscripts will sometimes be omitted if they are clear from the context. 
Expectation over random variables is denoted by $\E[\cdot]$. Sets are indicated by uppercase calligraphic letters $\mathcal{X}$. 
The complex zero-mean circularly symmetric Gaussian distribution of a vector is denoted by $\rvec{x}\sim\CN(0,\Lambda_{\vec{x}})$, where $\Lambda_{\vec{x}} = \E[\mathbf{x}\mathbf{x}^\dag]$ is the covariance matrix. All logarithms are in base two. 
In the context of matrix operations, $|\cdot|$,
$(\cdot)^T$, $(\cdot)^\dag$, and $\norm{.}$ represent the determinant, transpose, conjugate transpose, and Frobenius norm operators, respectively. 
\section{Channel Model and Mutual Information 
}
\subsection{Channel Model}
We consider transmission over $n$ channels in the presence of \ac{ASE} noise at the receiver. The channel is assumed to be constant during a transmission block of length $N$, and changes randomly and independently between the blocks. The assumption that the channel does not change within a block is consistent with the fact that the \ac{SOP} drifts at a much slower rate than typical transmission rates in optical links \cite{Karlsson:2000, Allen:2003} and is well established in optical communications literature \cite{Dar:2014_1,Dar_BlockWise:2014}.
The block length $N$ is chosen based on the application and the drift speed of the channel. 
In most optical communication systems, there is no feedback channel between the transmitter and the receiver. Therefore, the problem of input distribution optimization, corresponding to the case that the channel is known at the transmitter, is not investigated.
It is assumed that \ac{PMD} is negligible and all channel impairments, including nonlinearities and chromatic dispersion are ideally compensated, with the exception of polarization fluctuation and \ac{ASE} noise, which are modeled by a unitary matrix and \ac{AWGN}, respectively.

Since different blocks are independent, we just model the symbols in one transmission block in the following. The transmitted signal in each channel at time $\smash{k = 0,\ldots, N-1}$ is an $n$-dimensional random vector $\rvec{s}_k$ that takes on values from a set $\mathcal{S}$ of zero-mean constellation points. After filtering and resampling the received signals into one sample per symbol, the vector of received samples $\rvec{x}_k$ can be expressed as
\begin{equation}\label{eq:System_Model}
	\rvec{x}_k=\H\rvec{s}_k+\rvec{z}_k
\end{equation}
where the $n\times n$ matrix $\H$ represents a \ac{MIMO} channel and $\rvec{z}_k$ denotes the complex \ac{ASE} noise samples at time $k$, which is assumed to be $\CN(0,\sigmaz\In)$ and independent of $\rvec{s}_k$.
In the remainder of this paper, we will omit the time index $k$ explicitly for notational convenience. This is possible because the input and noise are independent and identically distributed (i.i.d.) over time.
We define the covariance matrix of the input vector as
\begin{equation}\label{eq:Input_CoVar}
	\Lambdax{s}=\E_{\rvec{s}}[\rvec{s}\rvec{s}^\dag].
\end{equation}

In order to maintain the generality of the results, they are given for an arbitrary number of channels $n$ whenever possible. Thus, the \ac{MIMO}-\ac{AWGN} channel \eqref{eq:System_Model} can describe a wide range of applications and impairments. However, for the purpose of \ac{DP} optical channel modeling, we are particularly interested in the special cases of $\smash{n=2}$ and $\H$ being unitary, denoted by $\Hu$.
\subsection{Mutual Information with Perfect Knowledge of Channel at the Receiver}
The conditional \ac{MI} between two random vectors $\rvec{s}$, $\rvec{x}$, when the channel $\smash{\Hb = \H}$ is given, is defined as \cite[Eq. 2.61]{CoverBook}
\begin{align}\label{eq:Conditional_MI}
    I(\rvec{s};\rvec{x}) = \E_{\rvec{s},\rvec{x}}\left[\logFrac{p(\rvec{x}|\rvec{s},\H)}{p(\rvec{x}|\H)}\right].
\end{align}
The capacity of this channel under an average power constraint is \cite{Shannon:1948}
\begin{align}
    C = \max\limits_{p(\rvec{s})} {I}(\rvec{s};\rvec{x}) \quad \text{s.t.} \quad \tr(\Lambdax{s})\le P,
\end{align}
where $P$ is the transmission power constraint. For the \ac{MIMO}-\ac{AWGN} channel in \eqref{eq:System_Model}, the channel law given $\H$ and $\vec{s}$ is characterized by the \ac{pdf}
\begin{equation}\label{eq:Channel_Law}
	p(\vec{x}|\vec{s},\H) = \dfrac{1}{\pin\sigma_z^{2n}}\expFrac{\norm{\vec{x}-\H\vec{s}}^2}{\sigmaz}.
\end{equation}
Then, the \ac{pdf} of the channel output can be calculated as
\begin{align}\label{eq:Channel_Output}
    p(\vec{x}|\H) = \E_{\rvec{s}}\left[p\right(\vec{x}|\rvec{s},\H)].
\end{align}
Given $\H$, the capacity-achieving distribution of the \ac{MIMO}-\ac{AWGN} channel law \eqref{eq:Channel_Law} is $\CN$ \cite{Telatar:1999}. Therefore, assuming $\CN(0,\Lambdax{s})$ inputs,
\begin{align}\label{eq:Channel_Output2}
    p(\vec{x}|\H) = \dfrac{1}{\pin|\Lambdax{x}|}\exp \left(-\vec{x}^{\dag}\Lambdax{x}^{-1}\vec{x}\right),
\end{align}
where
\begin{align}\label{eq:Lambda}
    \Lambdax{x} &= \E_{\rvec{x}}\left[\rvec{x}\rvec{x}^\dag\right]\nonumber\\
    &= \H\Lambdax{s}\H^\dag +\sigmaz\I_n
\end{align}
is the covariance matrix of the received samples when the channel is given.
The \ac{MI} of a general \ac{MIMO} system for a given channel is
\cite{Telatar:1999}
\begin{equation}\label{eq:MI_MIMO_Perfect_CSI}
	\begin{centering}
	    I(\rvec{s};\rvec{x}) = \log\left|\I_n +\H^{\dag} \H\Q\right|,
	\end{centering}	
\end{equation}
where 
\begin{align}\label{eq:Q}
    \Q = \Lambdax{s}/\sigmaz,
\end{align}
and hence $\smash{\tr(\Q)\le n\eta}$, where $\smash{\eta = P/(n\sigmaz)}$ is the \ac{SNR} of each channel. If the channel matrix $\H$ is confined to the set of unitary matrices (i.e., $\H = \Hu$), \eqref{eq:MI_MIMO_Perfect_CSI} gives
\begin{equation}\label{eq:MI_SOP_Perfect_CSI}
	I(\rvec{s};\rvec{x}) = \log\left|\I_n + \Q\right|.
\end{equation}
The capacity of a unitary \ac{MIMO}-\ac{AWGN} channel is the supremum of \eqref{eq:MI_SOP_Perfect_CSI} for all possible $\Q$. In general, $\Q$ needs to be optimized for each realization of the channel $\H$, if it is known at the transmitter; however, based on \eqref{eq:MI_SOP_Perfect_CSI}, the \ac{MI} is independent of $\H$. Since maximizing \eqref{eq:MI_SOP_Perfect_CSI} is equivalent to maximizing $|\smash{\In+\mathrm{Q}}|$ and $\Q$ is positive definite, the nondiagonal elements of $\Q$ must be zero, yielding $\Q$ to be a diagonal positive definite matrix. From the well-known theorem that the geometric mean is always upper-bounded by the arithmetic mean, it is straight-forward to show that uniform power distribution at the transmitter (i.e., $\smash{\Q = \eta \In}$) maximizes \eqref{eq:MI_SOP_Perfect_CSI}. Thus, the capacity of an $n$-dimensional unitary \ac{MIMO}-\ac{AWGN} channel, still assuming that the channel is perfectly known at the receiver, is 
\begin{equation}\label{eq:Cap_SOP_Perfect_CSI}
	\begin{centering}
	    C = n \log \left(1 + \eta \right).
	\end{centering}	
\end{equation}
The capacity of the \ac{DP} channel is given by simply setting $\smash{n=2}$ in \eqref{eq:Cap_SOP_Perfect_CSI}.

For a uniform input distribution over a given constellation $\mathcal{S}$, the integral in \eqref{eq:Channel_Output} must be replaced by a summation over all the constellation points, i.e.,
\begin{align}\label{eq:Output_PMF}
    p(\vec{x}|\H) = \sum_{\vec{s}\in \mathcal{S}} {P(\vec{s}) p(\vec{x}|\vec{s},\H)},
\end{align}
where $P(\vec{s})$ denotes the probability mass function of the input vector and $p(\vec{x}|\vec{s},\H)$ is defined in \eqref{eq:Channel_Law}. Using \eqref{eq:Conditional_MI} 
and \eqref{eq:Output_PMF}, the \ac{MI} of the channel for a uniformly distributed discrete input can be expressed as 
\begin{align}
    {I}(\rvec{s};\rvec{x})= \E_{\rvec{s},\rvec{x}}\left[\log\left( \frac{|\mathcal{S}|\exp(-\frac{\norm{\rvec{x}-\H\rvec{s}}^2}{\sigmaz})}{\underset{\vec{s}'\in\mathcal{S}}{\sum}\exp(-\frac{\norm{\rvec{x}-\H\vec{s}'}^2}{\sigmaz})}\right)\right],
\end{align}
where the expectations can be estimated numerically.
\section{the Mutual information in presence of Channel Estimation Error}\label{sec:ImperfectCSI}

We derived the capacity of the \ac{DP} channel with the assumption of perfect channel knowledge at the receiver in \eqref{eq:Cap_SOP_Perfect_CSI}.
In this section, we derive a lower bound on the \ac{MI} of the channel in the presence of an estimation error. As already shown in \cite{Telatar:1999}, when the channel is known, the capacity-achieving distribution is a zero-mean $\CN$ input with a power constraint. Thus, keeping the capacity-achieving distribution, i.e., $\smash{\rvec{s} \sim \CN(0,\Lambdax{s})}$ seems reasonable for the imperfectly estimated channel as well. 
To derive the \ac{AIR} $I_q$, which is a lower bound on the \ac{MI} \eqref{eq:Conditional_MI} between $\rvec{s}$ and $\rvec{x}$, the mismatched decoding inequality \cite{Agrell_Duality_2017}
\begin{equation}\label{eq:Miss_Decoding}
	I(\rvec{s};\rvec{x}) \ge I_q =  \E_{\rvec{x},\rvec{s}}\left[ \log\left(\frac{q(\rvec{x}|\rvec{s})}{q(\rvec{x})}\right)\right]	
\end{equation}
is used, where $q(\cdot)$ stands for the \ac{pdf} of an auxiliary channel. Note that this inequality holds for an arbitrary distribution of $q(\cdot)$. The mismatched channel law is here assumed to be  
\begin{equation}\label{eq:Missmatched_Channel_Law}
	q(\vec{x}|\vec{s}) =  p(\vec{x}|\vec{s},\Hhat),
\end{equation}
which is obtained by replacing $\H$ in \eqref{eq:Channel_Law} with the estimated channel $\Hhat$. This leads to
\begin{align}\label{eq:Missmatched_Channel_Output}
   q(\vec{x}) = p(\vec{x}|\Hhat).
\end{align}
Also, the covariance matrix of the auxiliary channel's output is obtained by replacing $\H$ with $\Hhat$ in \eqref{eq:Lambda}, i.e.,
\begin{align}\label{eq:Lambda_Hat}
\hat{\Lambda}_{\vec{x}} = \hat{\H}\Lambdax{s}\hat{\H}^\dag+ \sigmaz\In.
\end{align}
The average \ac{AIR} when the estimated channel is random can be written as
\begin{align}\label{eq:Random_Conditional_MI}
\bar{I}_q = \E_{\Hbhat}[I_q]
\end{align}

In the following, we first derive an \ac{AIR} for the \ac{MIMO}-\ac{AWGN} channel model with a fixed channel $\H$ and estimated channel $\Hhat$. Then, we extend the derived \ac{AIR} to $n$-dimensional unitary channels. By assuming a unitary estimate of the channel (i.e., $\smash{\Hhat \Hhat^\dag = \In}$), a tighter \ac{AIR} is derived. Finally, we use \eqref{eq:Random_Conditional_MI} to consider a random estimated channel $\Hbhat$.
\begin{theorem}\thlabel{Th:Lower_Bound}
Consider an arbitrary complex \ac{MIMO} channel matrix $\H$ and a fixed estimated channel matrix $\Hhat$. Defining $\smash{\mathrm{E} = \H-\Hhat}$, the \ac{AIR} is
\begin{align}
	I_q & =  \log \left|\In+\Hhat\Q\Hhat^\dag\right|-\frac{1}{\ln{2}}\tr \left(\Q \mathrm{E}^\dag \mathrm{E}\right) \nonumber\\
	&- \frac{1}{\ln{2}}\tr \left(\In - \Lambdax{x} \hat{\Lambda}_{\vec{x}}^{-1}\right),\label{eq:AIR_Theorem1}
\end{align}
where the definitions of $\Lambdax{x}$, $Q$, and $\LambdaHat$ can be found in \eqref{eq:Lambda},  \eqref{eq:Q}, and \eqref{eq:Lambda_Hat}, respectively.
\end{theorem}
\begin{IEEEproof}
One may rewrite \eqref{eq:Miss_Decoding} as
\begin{align}
	I_q &= \overbrace{-\E_{\rvec{x}}\left[\log q(\rvec{x})\right]}^ {A}
	+\overbrace{\E_{\rvec{s},\rvec{x}}\left[\log q(\rvec{x}|\rvec{s})\right]}^{B}.\label{eq:MI_Gen_Est_Miss_Decoding}
\end{align}
Using \eqref{eq:Channel_Output2} and \eqref{eq:Missmatched_Channel_Output}, given $\H$, the first term $A$ can be calculated as
\begin{align}
    A &= -\E_{\rvec{x}}\left[\log\left(\dfrac{1}{\pin|\hat{\Lambda}_{\vec{x}}|}\exp \left(-\rvec{x}^{\dag}\hat{\Lambda}_{\vec{x}}^{-1}\rvec{x}\right)\right)\right]\nonumber\\ 
	 &= \log \left(\pin|\hat{\Lambda}_{\vec{x}}|\right) +\frac{1}{\ln{2}} \E_{\rvec{x}}\left[\rvec{x}^\dag \hat{\Lambda}_{\vec{x}}^{-1}\rvec{x}\right] \nonumber \\
	&= \log \left(\pin|\hat{\Lambda}_{\vec{x}}|\right) +\frac{1}{\ln{2}}\tr\left(\Lambdax{x} \hat{\Lambda}_{\vec{x}}^{-1}\right),\label{eq:MI_Gen_Est_A}
\end{align}
where the last step follows the cyclic permutation rule of the trace. For the second term, $B$, when $\H$ is given, we use \eqref{eq:Channel_Law} and \eqref{eq:Missmatched_Channel_Law} to obtain
\begin{align}
    B =&~ \E_{\rvec{s},\rvec{x}}\left[\log\left( \dfrac{1}{\pin\sigmazn}\exp \left(-\frac{\norm{\rvec{x}-\Hhat\rvec{s}}^2}{\sigmaz} \right)\right)\right] \nonumber\\
	 =& -\log\left(\pin\sigmazn\right)-\frac{\E_{\rvec{z},\rvec{s}}\left[\norm{ (\H-\Hhat)\rvec{s}+\rvec{z}}^2\right]}{\sigmaz\ln{2}}\nonumber\\
	=& -\log\left(\pin\sigmazn\right)- \frac{\E_{\rvec{s}}\left[\norm{ \mathrm{E}\rvec{s}}^2\right]+\E_{\rvec{z}}\left[\norm{\rvec{z}}^2\right]}{\sigmaz\ln{2}}\label{eq:MI_Gen_Est_B_2}\\
	=& -\log\left(\pin\sigmazn\right)- \frac{\tr \left(\sigmaz\In+\mathrm{E}\Lambdax{s} \mathrm{E}^\dag \right)}{\sigmaz\ln{2}}\label{eq:MI_Gen_Est_B_3}\\
	=& -\log\left(\pin\sigmazn\right)- \frac{1}{\ln{2}}\tr \left(\In+\Q \mathrm{E}^\dag\mathrm{E} \right),\label{eq:MI_Gen_Est_B_4}
\end{align}
where \eqref{eq:MI_Gen_Est_B_2} follows from the fact that $\rvec{z}$ and $\rvec{s}$ are independent, and the cyclic permutation rule of the trace is used in \eqref{eq:MI_Gen_Est_B_3}. Finally, substituting \eqref{eq:MI_Gen_Est_A} and \eqref{eq:MI_Gen_Est_B_4} into \eqref{eq:MI_Gen_Est_Miss_Decoding} completes the proof.
\end{IEEEproof}

For the sake of keeping Theorem \ref{Th:Lower_Bound} as general as possible, no assumption is made about the transmitter knowledge of the estimated channel. However, in the context of the optical communications, we are more interested in the case that the transmitter has no channel knowledge. When the channel is not known at the transmitter, the only reasonable choice for $\Lambdax{s}$ is to apply uniform power distribution between the channels.
\begin{corollary}\thlabel{corollary:1}
Assume that the channel is unitary (i.e., $\smash{\H = \Hu}$) and that uniform power distribution takes place at the transmitter. Then the \ac{AIR} of a unitary channel can be written as 
    \begin{align}\label{eq:AIR_Corrollary1}
    	I_q &=  \log \left|\In+\eta\Hhat\Hhat^\dag\right|\nonumber-\frac{1}{\ln{2}}\tr
    	 \left(\eta\mathrm{E}^\dag \mathrm{E}\right) \\
    	 &-\frac{1}{\ln{2}}\tr\left( \In -(1+\eta)(\In+\eta\Hhat\Hhat^\dag)^{-1}\right).
    \end{align}
\end{corollary}
\begin{IEEEproof}
The uniform power distribution $\smash{\Lambdax{s} = P\In/n}$ yields $\smash{\Q = \eta\In}$. Knowing the channel is unitary (i.e., $\smash{\Hu\Hu^\dag = \In}$) and using \eqref{eq:Lambda} and \eqref{eq:Lambda_Hat}, we can write $\smash{\Lambdax{x} = \sigmaz(1+\eta)\In}$ and $\smash{\hat{\Lambda}_{\vec{x}}^{-1} = \sigma_z^{-2}(\In+\eta\Hhat\Hhat^\dag)^{-1}}$. Finally, substituting $\mathrm{Q}$, $\Lambdax{x}$, and $\hat{\Lambda}_{x}^{-1}$ in \eqref{eq:AIR_Theorem1} completes the proof.
\end{IEEEproof}
\vspace{1pt}
\begin{corollary}\thlabel{corollary:2}
Assume that the channel is a fixed unitary matrix $\Hu$ and that the estimated channel is an arbitrary random matrix $\smash{\Hbhat =  \Hu-\b{E}}$. Then with a uniform power distribution, the average \ac{AIR} is
    \begin{align}\label{eq:AIR_Rand_Est}
	    \bar{I}_q &= \E_{\Hbhat}\left[\log \left|\In+\eta\Hbhat\Hbhat^\dag\right|\right] -\frac{1}{\ln{2}}\tr \left(\eta\RE\right)\nonumber\\ 
	    &-\frac{1}{\ln{2}}\tr \left(\In -(1+\eta)\E_{\Hbhat}\left[(\In+\eta\Hbhat\Hbhat^\dag)^{-1}\right]\right),
    \end{align}
    where
    \begin{align}\label{eq:RE0}
    	\mathrm{R}_{\mathrm{E}}&= \E_{\b{E}}[\b{E}^\dag\b{E}].
    \end{align}
\end{corollary}
\begin{IEEEproof}
 By applying \eqref{eq:Random_Conditional_MI} to \eqref{eq:AIR_Corrollary1} the proof is complete.
\end{IEEEproof}
\begin{corollary}\thlabel{corollary:3}
If $\mathbf{E}$ has a spherically symmetric distribution, then \eqref{eq:AIR_Rand_Est} gives the same $\bar{I}_q$ for any $\Hu$. 
  \begin{IEEEproof}
Since $\Hu\Hu^\dag = \In$, we can write
\begin{align}
    \Hbhat\Hbhat^\dag &= (\Hu-\mathbf{E})(\Hu-\mathbf{E})^\dag \nonumber\\
    &= (\In-\mathbf{E}\Hu^\dag)(\In-\mathbf{E}\Hu^\dag)^\dag
\end{align}
and since $\mathbf{E}$ has a spherically symmetric distribution, it is invariant to rotation. Therefore, for any $\Hu$, $\mathbf{E}\Hu^\dag$ has the same distribution as $\mathbf{E}$ and $\Hbhat\Hbhat^\dag$ has the same distribution as $(\In-\mathbf{E})(\In-\mathbf{E})^\dag$. Thus, \eqref{eq:AIR_Rand_Est} yields the same \ac{AIR} independently of $\Hu$.
\end{IEEEproof}
\end{corollary}

We have not made any assumption on the estimation technique, so the derived lower bounds hold for an arbitrary estimator. It can be seen that \eqref{eq:AIR_Rand_Est} highly depends on the choice of estimation technique, so one can tighten the bound by choosing a suitable estimator. 
\vspace{1pt}

\begin{corollary}\thlabel{corollary:4}
For a complex unitary channel, a unitary estimated channel, and a uniform power distribution, the \ac{AIR} is
    \begin{align}\label{eq:AIR_Unitary_Est}
        \bar{I}_q &= n\log(1+\eta) - \frac{\eta}{\ln{2}}\tr (\RE).
     \end{align}
  \begin{IEEEproof}
    By applying $\smash{\Hbhat\Hbhat^\dag = \In}$ to \eqref{eq:AIR_Rand_Est} the proof is complete.
 \end{IEEEproof}
\end{corollary}

Note that \eqref{eq:AIR_Unitary_Est} is independent of $\H$, meaning that the \ac{AIR} is the same for any unitary channel. Interestingly, the unitary estimation of the channel removes two terms on the right-hand side of \eqref{eq:AIR_Rand_Est}, leading to a simpler bound. 

For uniformly distributed discrete inputs, the \ac{pdf} of the output of the auxiliary channel is
\begin{align}\label{eq:Pdf_Auxiliary_Output}
    q(\vec{x}) = \sum_{\vec{s}\in \mathcal{S}} {P(\vec{s}) q(\vec{x}|\vec{s})},
\end{align}
where $q(\vec{x}|\vec{s})$ is defined in \eqref{eq:Missmatched_Channel_Law}. Using \eqref{eq:Miss_Decoding}, \eqref{eq:Missmatched_Channel_Law}, and \eqref{eq:Pdf_Auxiliary_Output}, the average \ac{AIR} for a uniformly distributed discrete input can be expressed as 
\begin{align}\label{eq:AIR_Discrete}
    \bar{I}_q =  \E\left[\log\left( \frac{|\mathcal{S}|\exp(-\frac{\norm{\rvec{x}-\Hbhat\rvec{s}}^2}{\sigmaz})}{\underset{\vec{s}'\in\mathcal{S}}{\sum}\exp(-\frac{\norm{\rvec{x}-\Hbhat\vec{s}'}^2}{\sigmaz})}\right)\right],
\end{align}
where the expectation is over $\rvec{s}$, $\rvec{x}$, and $\Hbhat$, which can be estimated numerically.

\section{Channel Estimation}\label{sec:Channel_Estimation}
\begin{figure}
    \centering
    \includegraphics[width=.45\textwidth]{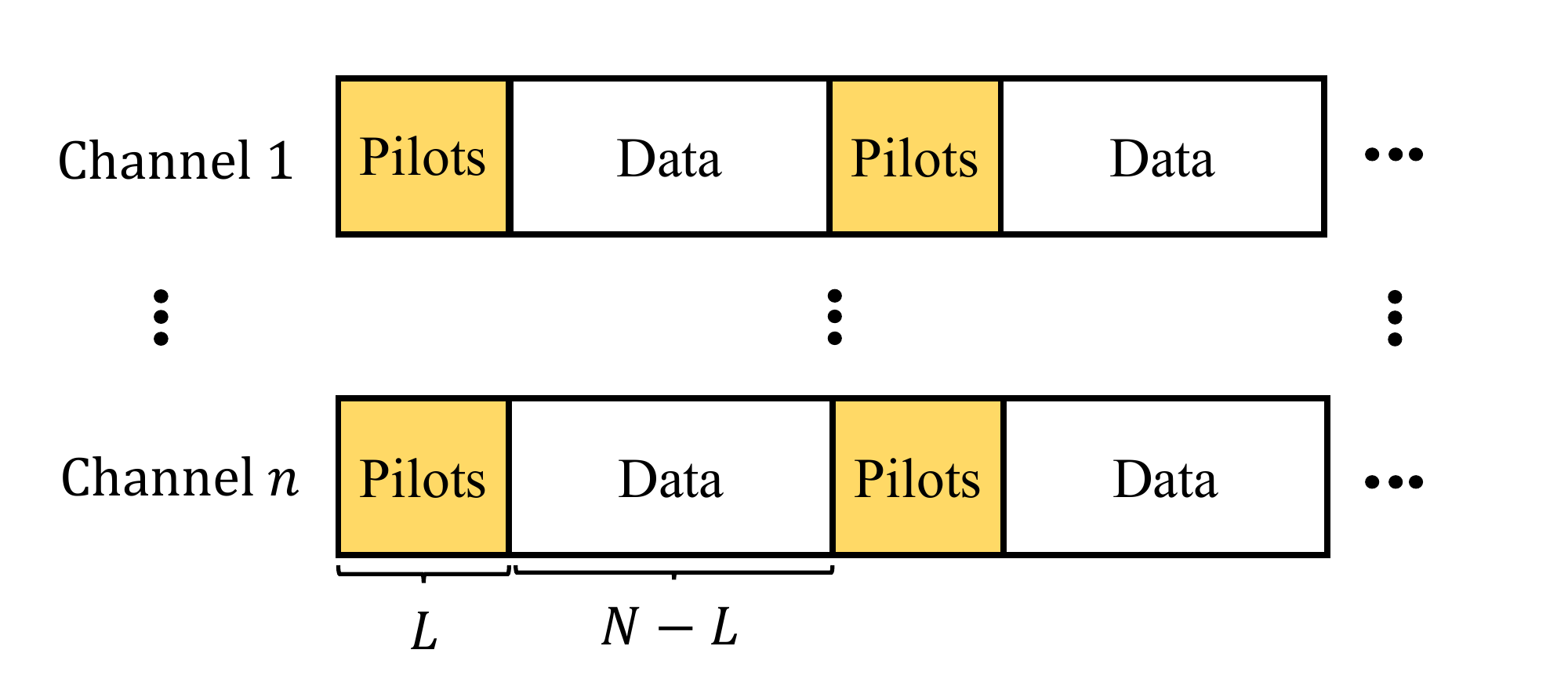}
    \caption{Transmission block model}
    \label{fig:pilot_distribution}
\end{figure}
In this section, first, the well-established \ac{LS} estimation algorithm is presented. Then, a unitary estimation method is proposed for unitary channels. As illustrated in Fig. \ref{fig:pilot_distribution}, to make data-aided channel estimation possible, the first $L$ symbols of the transmission block are pilot symbols $\smash{\mathrm{D} = [\vec{s}_0,...,\vec{s}_{L-1}]}$, which are assumed to be known at both transmitter and receiver. The optimal pilot assignment $\mathrm{D}$ should have the following properties \cite{Hassibi:2003}:
\begin{itemize}
    \item $\smash{L \ge n}$ 
    \item $\smash{\mathrm{D}\mathrm{D}^\dag = \frac{PL}{n} \In}$
\end{itemize}
The complex matrix $\smash{\b{X} = [\rvec{x}_0,...,\rvec{x}_{L-1}]}$ of received symbols is 
\begin{align}\label{eq:EstimationSystem_Model}
    \b{X}=\Hbu\mathrm{D}+\b{Z},
\end{align}
where $\smash{\b{Z} \sim \CN(0,\sigmaz\In)}$ is an $\smash{n\times L}$ matrix of i.i.d. noise samples and $\Hbu$ is an $\smash{n\times n}$ random unitary channel matrix, which is assumed to remain constant during a transmission block.

Note that the channel estimation problem can be translated to finding a certain number of independent real values, often regarded as the \ac{DOF} and denoted by $\nu$. 
\subsection{\ac{LS} Algorithm}\label{subsec:LS}
Conventional optical transmission systems often use \ac{LMS} for a real-time estimate of the channel, because \ac{LMS} tracks the channel change with each received symbol. However, in this paper it is assumed that the channel is constant during a transmission block and the \ac{LS} method is well adapted to block based transmissions. The \ac{LS} estimator estimates the channel by minimizing the squared error between the desired and received signal. For a \ac{MIMO}-\ac{AWGN} channel, the \ac{LS} optimization problem can be expressed as \cite{LS_Biguesh_2006}
\begin{align}\label{eq:LS_Estimation_Problem}
    \min\limits_{\Hbhat}  \norm{\b{X}-\hat{\Hb}\mathrm{D}}^2
\end{align}
 Knowing $\mathrm{D}$ and $\b{X}$, the solution of \eqref{eq:LS_Estimation_Problem} is \cite{LS_Biguesh_2006},\cite[Ch.~8]{Kay:1993} 
\begin{equation}
	\Hbhat_{\text{LS}} = \b{X}\mathrm{D}^\dag (\mathrm{D}\mathrm{D}^\dag)^{-1}
\end{equation}
However, the pilots are chosen in such a way that $\smash{\mathrm{D}\mathrm{D}^\dag = \frac{PL}{n}\In}$. Therefore, we can write
\begin{equation}
	\Hbhat_{\text{LS}} = \frac{n}{PL}\b{X}\mathrm{D}^\dag.
	\label{eq:LS_Estimation}
\end{equation} 
It can be shown that $\mathrm{R_{E}}$ for the LS algorithm is \cite{LS_Biguesh_2006}
\begin{align}\label{eq:RE_LS}
    \mathrm{R_{E}} = \E_{\mathbf{E}}\left[\mathrm{\mathbf{E}}^\dag\mathrm{\mathbf{E}}\right]= \frac{n\sigmaz}{PL}\In = \frac{1}{\eta L}\In
\end{align}
showing that the estimation error of the \ac{LS} algorithm is inversely proportional to the \ac{SNR} and the pilot length $L$. 

It is worth mentioning that the \ac{LS} algorithm has to find $\nu = 2n^2$ independent real values to make an estimate.
\subsection{Kabsch Algorithm}\label{subsec:Kabsch}
The problem with both blind estimation algorithms (e.g., \ac{CMA}, \ac{RDE}, and \ac{MMA}) and pilot-aided estimation algorithms (e.g., \ac{LS} and \ac{LMS}) is that their optimization problems are not adopted for unitary channel estimation, making their solutions suboptimal if $\Hb$ is known to be unitary. Thus, in this part, we apply the unitary constraint of the channel to the estimation problem of \eqref{eq:LS_Estimation_Problem} and write
\begin{align}
     \min\limits_{\Hbhat} \norm{\b{X}-\Hbhat\mathrm{D}}^2 \quad\text{s.t.} \quad \Hbhat\Hbhat^\dag  = \In\label{eq:ULS_Opt_Problem}.
\end{align}
The optimal solution to this problem is given by the Kabsch algorithm \cite{Kabsch:1976} as
\begin{align}
    	\Hbhat_{\text{Kabsch}} = \b{UV}^\dag,\label{eq:ULS_Opt_Solution}
\end{align}
where $\smash{\b{U}\Sigma \b{V}^\dag}$ is the singular value decomposition function of $\b{X}\mathrm{D}^\dag$. As can be seen, unlike the blind conventional estimators, Kabsch is independent of the modulation format of the transmission. The Kabsch algorithm was proposed for optical communication by Louchet \textit{et al.} \cite{Louchet:2014} as a blind polarization tracking algorithm, where decision-directed symbols were used instead of pilots.

In contrast to \ac{LS}, no analytical result is known for $\mathrm{R_E}$ of the Kabsch algorithm. Although we cannot analytically prove it, we can make an intuitive prediction by considering that $\smash{\nu = n^2}$ for a unitary estimation of the channel while for a general estimation of the channel $\smash{\nu = 2n^2}$. Since the channel estimation problem is equivalent to finding $\nu$ independent real-valued quantities, one can predict that the estimation error of Kabsch would be half of \ac{LS}. More interestingly, for special unitary channels where $\smash{|\Hu| = 1}$, the \ac{DOF} is $\smash{n^2-1}$ and the gain by the unitary algorithm can be even higher; however, this gain vanishes for large $n$. 

Note that in the case of the \ac{DP} channel, the singular value decomposition is deployed only on a two-by-two matrix, making it less computationally complex than for higher $n$. 

\section{Numerical Results}\label{sec:V}
In this section, through Monte Carlo trials, the \ac{AIR} of the \ac{DP} channel (i.e., $\smash{n=2}$) for the estimation algorithms detailed in Section \ref{sec:Channel_Estimation} is computed. While \ac{AIR}s are derived for a fixed channel matrix $\H$, the estimated channel $\Hbhat$ is dependent on each realization of the channel. 
A deterministic sequence of $L$ \ac{QPSK} symbols is selected to satisfy the pilot conditions detailed in Section \ref{sec:Channel_Estimation}. 
\begin{figure}
    \centering
    \includegraphics[width=.45\textwidth]{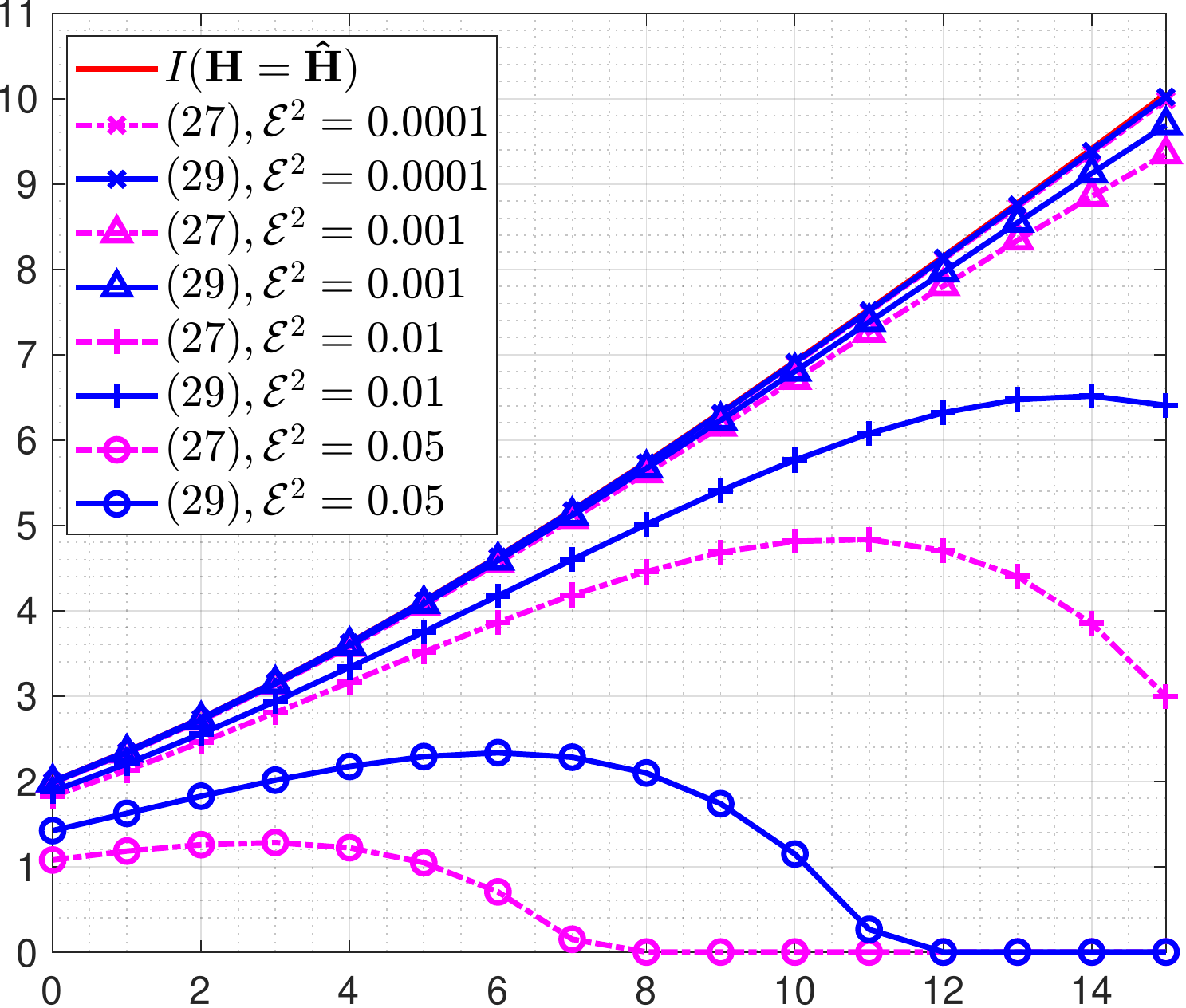}
    \put(-120,-10){\small{$\eta$ (dB)}}
    \put(-240,35){\rotatebox{90}{\small{Information Rate (bits/symbol)}}} 
    \caption{The average \ac{AIR} $\bar{I}_q$ of the \ac{DP} channel as a function of \ac{SNR} $\eta$ and estimation error per \ac{DOF} $\mathcal{E}^2$.}
    \label{fig:LB_Unitary_vs_General}
\end{figure}

Numerical results verify that the estimation error of the Kabsch algorithm completely follows our prediction in Sec.~\ref{subsec:Kabsch}. Thus, to perform a fair comparison between the unitary and nonunitary estimators, we define a new parameter called estimation error per \ac{DOF} as $\mathcal{E}^2 = \tr{\mathrm{R}_\mathrm{E}}/(n\nu)$. Note that for a general nonunitary estimator $\nu = 2n^2$ and for a general unitary estimator $\nu = n^2$.

Using $\CN$ inputs, Fig. \ref{fig:LB_Unitary_vs_General} shows the Monte-Carlo averaged \ac{AIR} $\bar{I}_q$ of the \ac{DP} channel according to \eqref{eq:AIR_Rand_Est} and \eqref{eq:AIR_Unitary_Est} for $\smash{n=2}$. Note that based on \thref{corollary:3}, for an estimation error $\mathbf{E}$ with a spherically symmetric distribution, \eqref{eq:AIR_Rand_Est} is independent of $\Hu$. For the blue curves, we have assumed that $\mathbf{E}$ is distributed according to $\CN(0,\mathrm{R_E})$ with $\smash{\mathrm{R}_{\mathrm{E}}= 2n^2\mathcal{E}^2\I_2}$. 
For the dashed magenta curves, $\Hbhat$ is unitary and $\smash{\mathrm{R}_{\mathrm{E}}= n^2\mathcal{E}^2}\In$. The results illustrate that when we have the same $\mathcal{E}^2$, the unitary estimate of the channel leads to a higher \ac{AIR}. Besides, it can be seen that for a constant estimation error per \ac{DOF} $\mathcal{E}^2$, the average \ac{AIR} reaches a maximum for a specific \ac{SNR} and then decreases. This can be justified according to the right-hand side of \eqref{eq:AIR_Rand_Est} and \eqref{eq:AIR_Unitary_Est}, where the first term increases in a logarithmic manner with respect to \ac{SNR}, but the second term decreases linearly with \ac{SNR}. Therefore there is an optimum \ac{SNR} that maximizes the AIR. This behavior comes from the assumption of a constant $\mathcal{E}^2$ regardless of the \ac{SNR}, which, as we shall see next, is not fully realistic. 

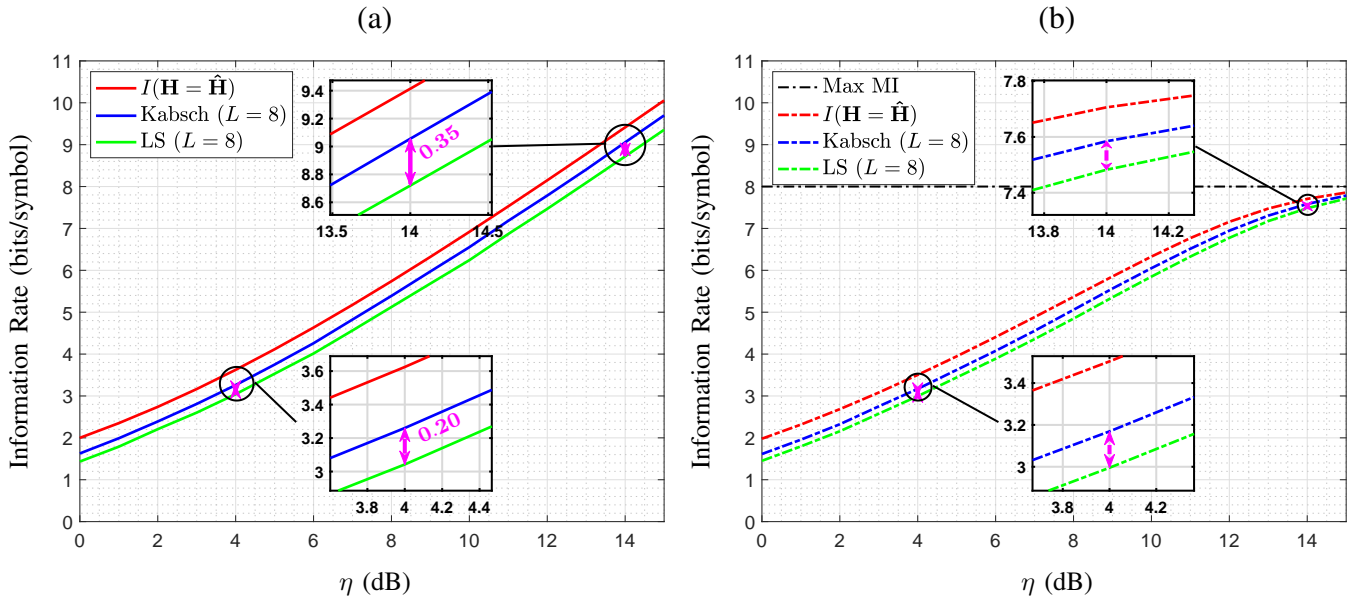
\begin{figure*}
	\centering
	\begin{subfigure}{.5\textwidth}
	 	\begin{tikzpicture}
		\node at (0,0){\includegraphics[width=.90\textwidth]{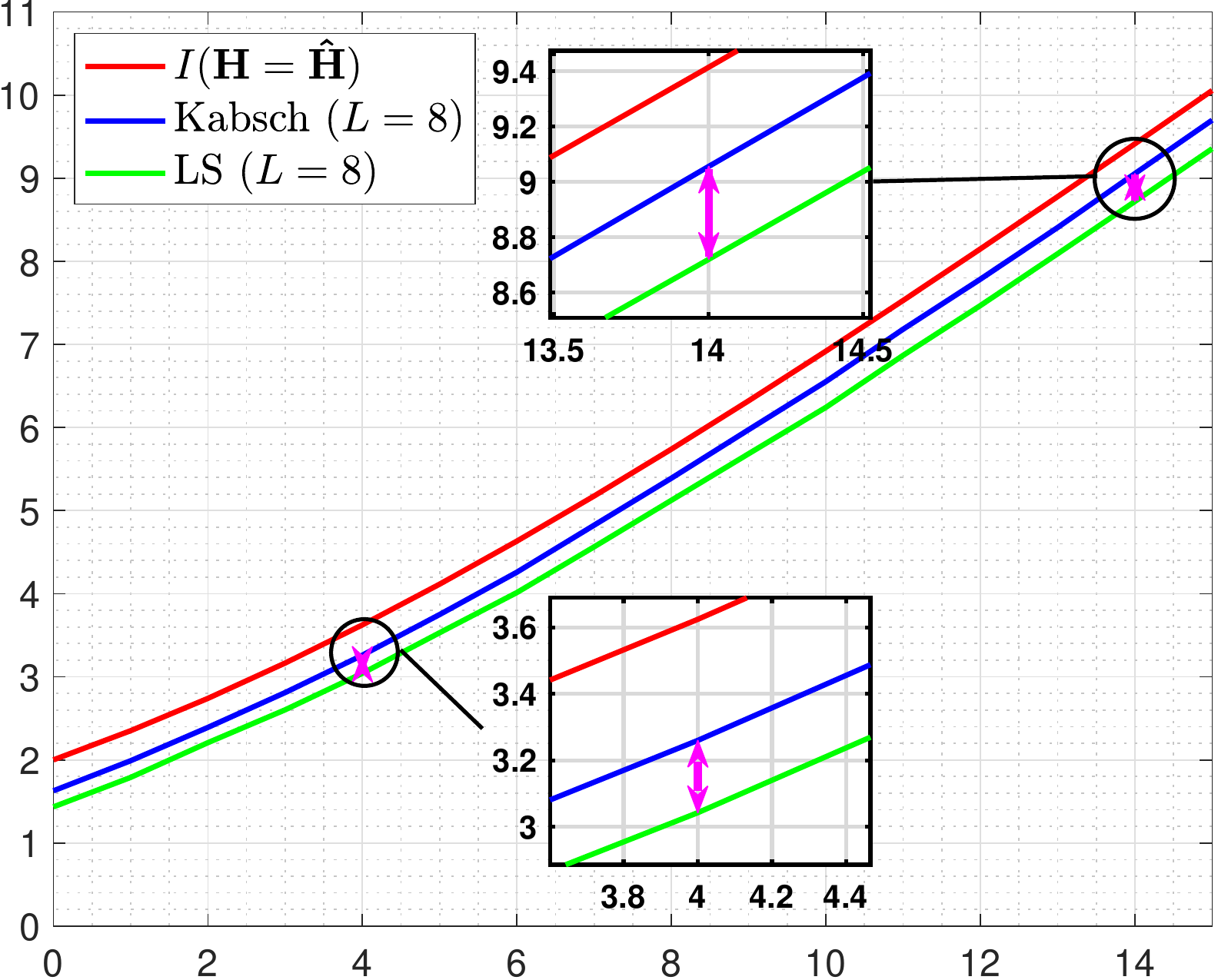}};
		\node at (0.2,3.75){{(a)}};
        \node at(0.2,-3.75){\small{$\eta$ (dB)}};
        \node at(-4.5,0){\rotatebox{90}{\small{Information Rate (bits/symbol)}}};
		\node at (1.10,2.15){\rotatebox{35}{\scriptsize{\textcolor{myMagneta}{$\mathbf{0.35}$\textbf{ }}}}};
		\node at (1.10,-1.65){\rotatebox{25}{\scriptsize{\textcolor{myMagneta}{$\mathbf{0.20}$ \textbf{}}}}};
		\end{tikzpicture}
	\end{subfigure}%
	\begin{subfigure}{.5\textwidth}
		\begin{tikzpicture}
		\node at (0,0){\includegraphics[width=.90\textwidth]{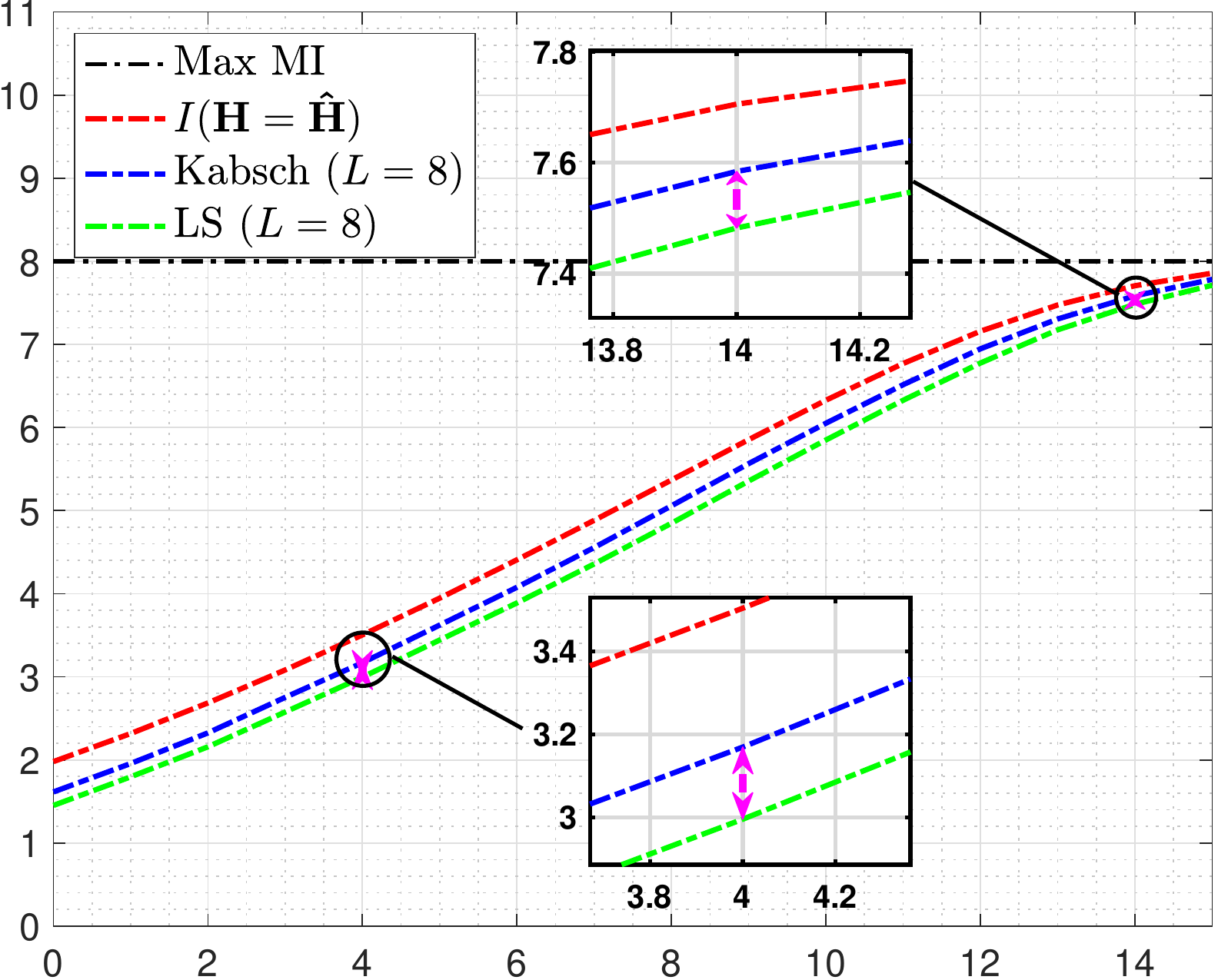}};
	    \node at (0.2,3.75){{(b)}};
        \node at(0.2,-3.75){\small{$\eta$ (dB)}};
        \node at(-4.5,0){\rotatebox{90}{\small{Information Rate (bits/symbol)}}};
		\end{tikzpicture}
	\end{subfigure}%
	\caption{\label{fig:AIR_Kabsch_LS}The average \ac{AIR} $\bar{I}_q$ when $\smash{L = 8}$ \ac{DP} pilot symbols are used. (a) Using $\CN$ inputs where for \ac{LS} and Kabsch, $\bar{I}_q$ is according to \eqref{eq:AIR_Rand_Est} and \eqref{eq:AIR_Unitary_Est}, respectively. (b) Using uniformly distributed DP-$16$-\ac{QAM} inputs where $\bar{I}_q$ for both \ac{LS} and Kabsch is according to \eqref{eq:AIR_Discrete}.}
\end{figure*}
Fig. \ref{fig:AIR_Kabsch_LS}(a) presents the \ac{AIR} of \ac{LS} and Kabsch when $\CN$ inputs are used. The solid red line indicates the receiver perfectly knows the channel. It can be concluded that with $\smash{L = 8}$, Kabsch surpasses \ac{LS} throughout the range of considered \ac{SNR}s. More specifically, at $4$ and $14$ dB \ac{SNR}s (see insets), Kabsch has at least $0.2$ and $0.35$ bits per symbol higher \ac{AIR}, respectively. Additionally, it is clear from the results that \ac{LS} is upper-bounded by Kabsch, and increasing the \ac{SNR} cannot fill the gap. This behavior can be justified because, unlike \ac{LS}, Kabsch guarantees a unitary estimation of the channel leading to a lower estimation error. Moreover, as the \ac{SNR} increases, the gap between Kabsch and LS and the actual channel MI is almost constant. Since the theoretical gap is $(\eta/\ln{2}) \tr(\mathrm{R_E})$ according to \thref{corollary:4}, we conclude that the error covariance $\RE$ of Kabsch is inversely proportional to the \ac{SNR}, i.e., $\smash{\mathrm{R}_{\mathrm{E}} \propto {1}/{\eta}}$. Unlike Fig. \ref{fig:LB_Unitary_vs_General}, the AIR bounds are monotonically increasing with SNR which is due to the fact that the estimation error is decreasing with SNR. 

A comparison between \ac{LS} and Kabsch with \ac{DP}-$16$-\ac{QAM} inputs is provided in Fig. \ref{fig:AIR_Kabsch_LS}(b). The \ac{MI} when the channel is perfectly known at the receiver is marked by the solid red line. Evidently, Kabsch outperforms LS throughout the considered range of SNRs. 
The results also support the fact that Kabsch upper-bounds \ac{LS} for various inputs, which completely agrees with Fig. \ref{fig:LB_Unitary_vs_General}, where the unitary estimation of the channel leads to a higher \ac{AIR}.

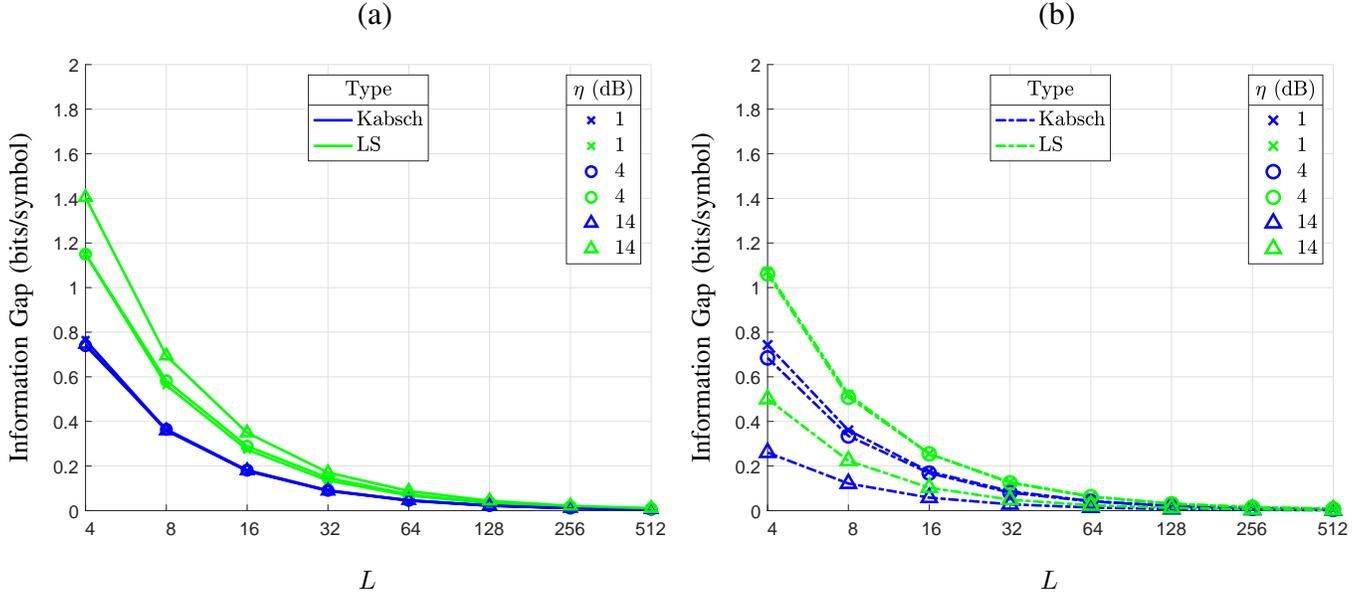
\begin{figure*}
	\centering
	\begin{subfigure}{.5\textwidth}
	 	\begin{tikzpicture}
		\node at (0,0){\includegraphics[width=.90\textwidth]{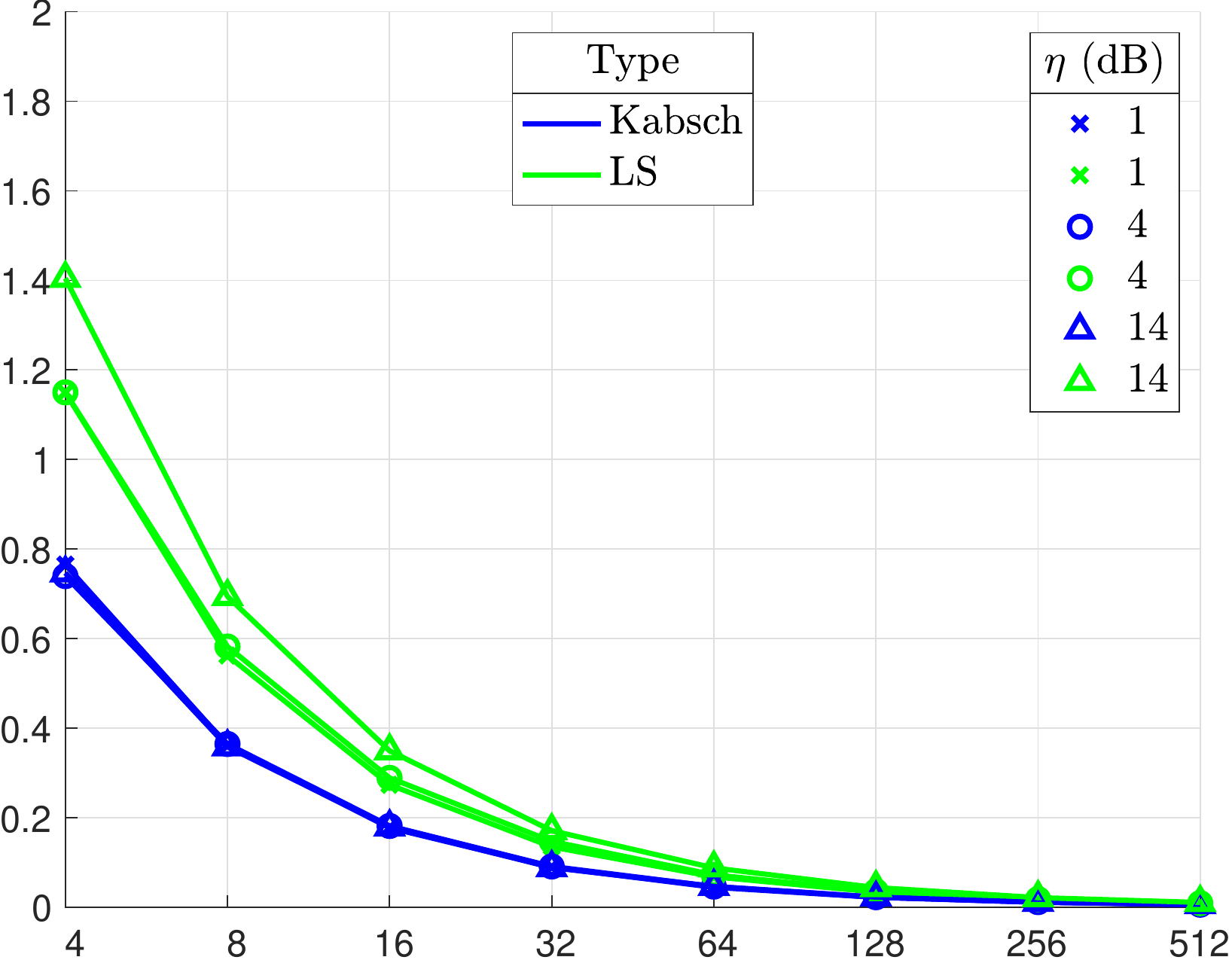}};
	    \node at (0.2,3.75){{(a)}}; 
        \node at(0.1,-3.75){\small{$L$}}; 
        \node at(-4.5,0){\rotatebox{90}{\small{Information Gap (bits/symbol)}}}; 
		\end{tikzpicture}
	\end{subfigure}%
	\begin{subfigure}{.5\textwidth}
		\begin{tikzpicture}
		\node at (0,0){\includegraphics[width=.90\textwidth]{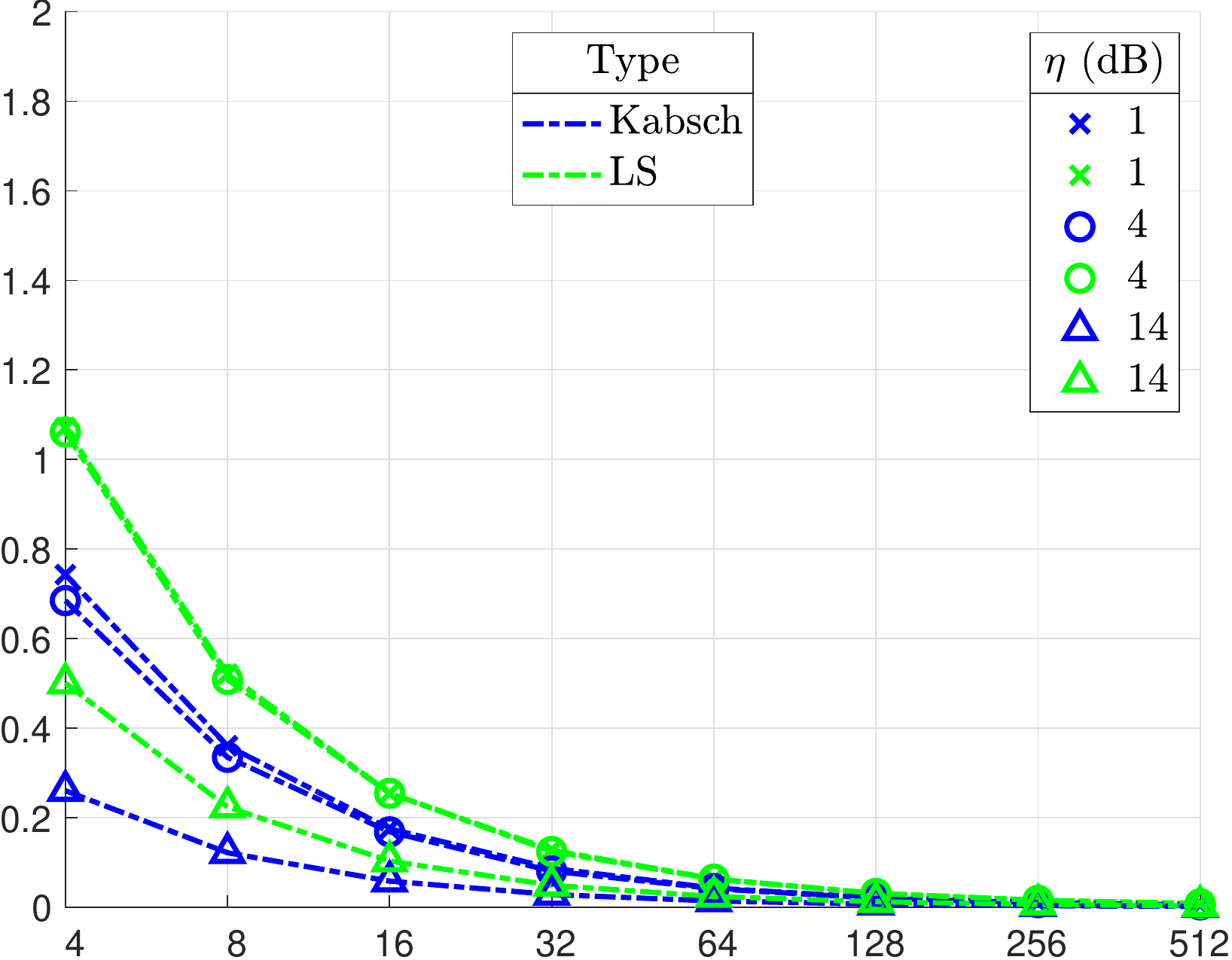}};
		\node at (0.2,3.75){{(b)}};
        \node at(0.1,-3.75){\small{$L$}};
        \node at(-4.5,0){\rotatebox{90}{\small{Information Gap (bits/symbol)}}}; 
		\end{tikzpicture}
	\end{subfigure}%
	\caption{\label{fig:AIR_GAP}The average information gap $I-\bar{I}_q$ for a range of pilot lengths. (a) $\CN$ inputs where $\bar{I}_q$ for \ac{LS} and Kabsch algorithms is according to \eqref{eq:AIR_Rand_Est} and \eqref{eq:AIR_Unitary_Est}, respectively. (b) Uniformly distributed \ac{DP}-$16$-\ac{QAM} inputs where $\bar{I}_q$ for both \ac{LS} and Kabsch is according to \eqref{eq:AIR_Discrete}.}
\end{figure*}

Fig. \ref{fig:AIR_GAP} displays the information gap between the \ac{AIR}s and the capacity of the \ac{DP} channel, when (a) $\CN$ inputs and (b) \ac{DP}-$16$-\ac{QAM} inputs are used. The overall dominance of Kabsch throughout the range of considered \ac{SNR}s can be easily verified. Evidently, it is beneficial to use higher pilot numbers at low \ac{SNR}s. Given the fact that $\smash{\mathrm{R}_{\mathrm{E}} \propto {1}/({L\eta})}$ for \ac{LS} \eqref{eq:RE_LS} and that Kabsch in Fig. \ref{fig:AIR_GAP} follows the same trend with respect to $L$, we can empirically conclude that $\mathrm{R_E}$ of Kabsch is also inversely proportional to the pilot length and the \ac{SNR}.

It is beneficial to use Kabsch to limit the rate loss due to the pilots. For example, a reasonable performance is achieved by only a block of $\smash{L=16}$ pilot symbols, which is relatively small compared to the transmission block size in optical communication systems, implying that the rate loss is negligible. For instance, for a system operating at a rate of $28$ Gbaud, even if the channel remains constant for one microsecond (i.e., the \ac{SOP} drift time is one microsecond), it corresponds to a block length $N$ of at least $28,000$ symbols and the rate loss due to $16$ pilots is negligible.

\section{Conclusion}
The capacity of unitary MIMO-AWGN channels and specifically \ac{DP} was investigated. With perfect channel knowledge, the capacity of unitary channels is the same as for the regular \ac{AWGN} channel. An \ac{AIR} with imperfect channel knowledge was derived and showed that the \ac{AIR} is highly dependent on the estimation algorithm. In the case of unitary channels, higher \acp{AIR} are obtained with a unitary estimation of the channel. The bounds are derived for any $\smash{n\ge 2}$ dimensions, meaning that the results can be applied to other optical channels. In particular, \thref{Th:Lower_Bound} can be directly applied to space-division multiplexed channels which are impaired with polarization- and mode-dependent loss. In contrast to the conventional estimation algorithms, the proposed data-aided Kabsch algorithm ensures a unitary estimate of the channel. Numerical results showed that for various input distributions, Kabsch outperforms \ac{LS} in terms of \ac{AIR}. 
Also, like \ac{LS}, Kabsch can perform very well with only a few pilot symbols, making the transmission rate loss due to the pilots negligible.
\ifCLASSOPTIONcaptionsoff
  \newpage
\fi



%

\bibliography{References}

%







\end{document}